\documentclass[12pt,preprint]{aastex}

\shorttitle{Intra-Day Variability of Sagittarius A$^*$ at 3 
Millimeters}
\shortauthors{Mauerhan et al.}

\begin{document}
\title{Intra-day Variability of Sagittarius A$^*$ at 3 Millimeters}
\author{Jon C. Mauerhan\altaffilmark{1}, Mark Morris\altaffilmark{1}, Fabian Walter\altaffilmark{2}, and Frederick K. Baganoff\altaffilmark{3}}
\altaffiltext{1}{Department of Physics \& Astronomy, University of California, Los Angeles, Los Angeles, CA 90095; mauerhan@astro.ucla.edu, morris@astro.ucla.edu} 
\altaffiltext{2}{Max Planck Institut f\"ur Astronomie, K\"onigstuhl 17, 69117 Heidelberg, Germany; walter@mpia.de}
\altaffiltext{3}{Center for Space Research, Massachusetts Institute of Technology, Cambridge MA 02139; 
fkb@space.mit.edu}

\begin{abstract}
We report observations and analysis of flux monitoring of Sagittarius 
A$^*$ at 3-mm wavelength using the OVRO millimeter interferometer over a period of 
eight days (2002 May 23--30). Frequent phase and flux 
referencing (every 5 minutes) with the nearby calibrator source J1744$-$312 was employed 
to control for instrumental and atmospheric effects. Time variations are 
sought by computing and subtracting, from each visibility in the 
database, an average visibility obtained from all the data acquired in our 
monitoring program having similar $\mathit{uv}$ spacings. This removes the 
confusing effects of baseline-dependent, correlated flux interference 
caused by the static, thermal emission from the extended source Sgr A West. 
Few-day variations up to $\sim{20\%}$ and intra-day variability of 
$\sim{20\%}$ and in some cases up to $\sim{40\%}$ on few-hour time scales emerge 
from the differenced data on SgrA$^*$. Power spectra of the residuals 
indicate the presence of hourly variations on all but two of the eight days. 
Monte Carlo simulation of red-noise light curves indicates that the hourly variations are well described by a red-noise power spectrum with $P(f)\propto{f^{-1}}$. Of particular interest is a $\sim{2.5}$ hour variation seen prominently 
on two consecutive days. An average power spectrum from all eight days 
of data reveals noteworthy power on this time scale. There is some 
indication that few-hour variations are more pronounced on days when the 
average daily flux is highest. We briefly discuss the possibility that 
these few-hour variations are due to the dynamical modulation of accreting 
gas around the central supermassive black hole, as well as the 
implications for the structure of the SgrA$^*$ photosphere at 3 mm. Finally, 
these data have enabled us to produce a high 
sensitivity 3-mm map of the extended thermal emission surrounding SgrA$^*$.
\end{abstract}


\section{INTRODUCTION}

It is now widely accepted that the compact radio source SgrA$^*$ is 
associated with the presence of a black hole of mass $\sim4\times10^6$ 
M$_\odot$ located at the dynamical center of the Galaxy (Ghez et al. 2003, 
Sch\"odel et al. 2003). Although SgrA$^*$ represents a relatively 
faint galactic nucleus, with L$\sim{10^{-8.5}}$ L$_{Edd}$ (Melia \& Falcke 
2001), ongoing detections of variability and flares at radio, 
sub-millimeter, infrared, and X-ray wavelengths indicate that this is a dynamic 
object powered by accretion.  At centimeter and millimeter wavelengths, 
persistent variability of SgrA$^*$  has been observed to occur on time 
scales of weeks with $\sim$3 active episodes per year, each lasting 
about a month (Herrnstein et al. 2004, Zhao et al. 2003, Falcke 1999). 
Furthermore, there appears to be a time-lagged correlation between the centimeter 
and millimeter fluxes, with variations at shorter wavelengths appearing 
stronger in amplitude and peaking out before the longer wavelengths 
(Zhao et al. 2003). This is consistent with a model in which 
the higher frequency emission originates from smaller spatial scales, 
closer to the event horizon of the black hole. Short-term variability of 
SgrA$^*$ at near-infrared (Ghez et al. 2004, Genzel et al. 2003) and 
X-ray (Baganoff 2001, 2003) wavelengths has also been observed recently. 
Eckart et al. (2004) have reported the first simultaneous detection of 
an X-ray and NIR flare from Sgr A$^*$. Clearly, characterizing the 
variability time scales at different wavelengths will help elucidate the structure and dynamics of this object.  

Also of interest are the effects of radiative transfer on our ability 
to detect variability on hourly time scales in the radio/sub-millimeter 
regime.  The size of SgrA$^*$ is dominated by scattering from the 
turbulent interstellar plasma along the line of sight for $\lambda \gtrsim $ 
3 mm and perhaps also by the optical depth of the plasma flow 
surrounding the black hole, resulting in a stratification of Sgr A$^*$'s 
appearance as a function of wavelength (Bower et al. 2004). Owing to the latter, 
the shortest observed time scales for flux variations at millimeter 
and sub-millimeter wavelengths may reveal important information about 
the structure of the SgrA$^*$ photosphere, since the detectability of 
flux variations of a given region will depend on whether the variable 
flux emanates from within or outside the $\tau\sim$1 surface.

One complication of the interpretation of measurements by 
interferometer arrays tracking SgrA$^*$ is the non-negligible flux component 
contributed by the extended thermal source SgrA West, which is sampled in a 
time varying, baseline-dependent fashion during the observing track.  
Atmospheric effects must also be controlled. To discriminate true variability 
from possible atmospheric fluctuations, one must interweave frequent 
measurements of a nearby phase calibrator with measurements of SgrA$^*$, 
especially for interferometers in the northern hemisphere, where 
SgrA$^*$ is observed at low elevations.  Perhaps for these reasons, 
low-amplitude variability on hourly and sub-hourly time scales has seldom been 
claimed with any confidence. One exception has been the report of a 
 30$\%$ intra-day flux increase and a 400$\%$ flare with $\sim$2-fold increase
 timescale of 1.5 hours at 2 mm by Miyazaki et al. (2004) and a 10$\%$ increase
 at 15 GHz over a 2 hour period reported by Bower et al. (2002). 

In this letter, we present the results of eight successive days of 
observation of SgrA$^*$ at 3 mm using Caltech's Owens Valley Radio Observatory's
millimeter interferometer (OVRO). This work was carried out as part of 
a multi-wavelength campaign, including the \emph{Chandra X-ray 
Observatory}, with the goal of observing radio counterparts to X-ray flares, 
although none occurred while OVRO was observing SgrA$^*$ 
(Baganoff et al. 2005). Nonetheless, the eight days of coverage analyzed 
here have enabled us to detect persistent, low-amplitude variations of Sgr 
A$^*$ on time scales of a few hours and to explore the nature of this 
intra-day variability.

\section{OVRO OBSERVATIONS}
The Galactic Center (SgrA$^*$) was observed with the OVRO millimeter 
interferometer in the L configuration at elevations $>15^\circ$ on each 
of the eight successive days of the multi-wavelength observing campaign 
(2002 May 23--30). The observing track each day lasted about 6 hours. 
Data were recorded in both sidebands using 2 IFs (offset from the local 
oscillator by $\pm$ 1.5, and $\pm$ 3 GHz), resulting in a total continuum 
bandwidth of 4 GHz ($\nu_{LO}$=101.8 GHz during the first 3 days and 
$\nu_{LO}$=96.3 GHz for the remaining 5 days).

After measuring the fluxes of $\sim10$ potential calibrator sources 
near Sgr A$^*$ in the week prior to the observing run, we chose 
J1744$-$312, separated from SgrA$^*$ by only 2.29$^\circ$. Its flux of 0.9 Jy at 
100 GHz is adequate for accurate amplitude and phase calibration and we 
observed it for 3 minutes after every 5 minute integration on SgrA$^*$.  
During each track, absolute flux calibration was done using the planets 
Uranus and Neptune, and was checked by observing secondary 
calibrators (3C273, 3C345, 3C454.3, 3C84, NRAO 530). 

The data were edited and calibrated with the Caltech millimeter array reduction package MMA. 
During most of the daily tracks the weather was exceptional for that time of the year. 
All eight days of data were used to construct our model of the source (Figure 1) 
and were phase-only self-calibrated in AIPS. To mimimize the inclusion of bad data, 
we deleted all visibilities with coherence $<$0.8.

\section{ANALYSIS AND RESULTS}
 In order to discriminate between intrinsic flux variations in the 
central point source and flux variations due to changes in the spatial 
frequencies being sampled from SgrA West (Figure 1) as the interferometer 
tracks the source across the sky, it is preferable to analyze the data 
in the $\mathit{uv}$ plane.  In removing the flux contribution of SgrA 
West, we make use of the fact that all visibilities in our data set are 
the vector sum of components from SgrA$^*$ and SgrA West. For every 
visibility $\mathit{V_{i}}$, we average the amplitudes and phases of all other visibilities $\mathit{V_{j}}$ within an 
averaging kernel of radius $r_{i}$ about $\mathit{V_{i}}$. The amplitudes and phases are weighted inversely by their distance 
$\mathit{dr_{ij}}$, in the $\mathit{uv}$ plane, from the central visibility $\mathit{V_{i}}$ and by their 
intrinsic weight $\mathit{w_{j}=\sigma_{j}^{-2}}$ due to system noise 
(we also experimented with gaussian kernels, although the resulting 
difference was negligible). The total weight for each visibility is then given by 
$\mathit{W_{j}=w_{j}*(1-dr_{ij}/{r_{i}})}$. We have constructed the size of $r_{i}$ to 
vary linearly with respect to distance from the $\mathit{uv}$ origin. This 
accounts for the decreasing $\mathit{uv}$ sampling density for the 
longer baselines, resulting in each averaging kernel having a comparable 
number of visibilities contributing to the average and thus uniform 
statistics throughout the entire $\mathit{uv}$ plane. By choosing an 
appropriate scaling for $r_{i}$, we limit the intra-kernel phase 
variations due to source structure to $\lesssim{10^\circ}$, ensuring that all 
intra-kernel visibilities sample the same extended structure component 
of the source. The final kernel used has a radius range of 0.3--3.0 
k$\lambda$ for the shortest ($\sim$4 k$\lambda$) to longest baselines 
($\sim$40 k$\lambda$). The average phase and amplitude of all remaining visibilities 
within the kernel are then used to construct a weighted average 
visibility $\mathit{V_{ave,i}=\sum_{j}^{} W_{j}V_{j}/\sum_{j}^{} W_{j}}$ that 
represents the global average of the data set (all eight days of data) in 
that $\mathit{uv}$ sector. We then compute a residual visibility 
$\mathit{V_{res,i}}$ at each point by subtracting $V_{ave,i}$ from $V_{i}$ and 
taking the real part of the difference: 
$\mathit{V_{res,i}}=\mathbb{R}e(V_{i}-V_{ave,i})$. Thus, $\mathit{V_{res,i}}$ represents true variability 
of SgrA$^*$, since the constant contribution from SgrA West and the average value of SgrA$^*$ have been 
subtracted out. Any variations due to atmospheric refraction would be 
evidenced in the imaginary part of the residuals of SgrA$^*$ (which have 
been confirmed to be flat) and in the real residuals of the calibrator 
J1744$-$312, which was subjected to the same treatment. For flux reference, we assume that the calibrator flux is constant, and fix its value at 5 minute intervals and time-average the data using 10 minute bins. Thus, we compare the real residual, intra-day fluxes of SgrA$^*$ with those from the 
calibrator J1744$-$312 for each day of the campaign.

Figure 2 shows the day-averaged flux residuals of SgrA$^*$ on each day of the campaign. Figure 3 exhibits the intra-day variations, showing the 
real components of the residual visibility vectors of SgrA$^*$ (column 
1) and J1744$-$312 (column 2) as well as the Lomb-Scargle power spectrum of 
the residuals (solid curve, column 3). Each residual light curve in Figure 3 is the average of the residuals from each separate baseline . All baselines exhibit residuals that are consistent with the average; i.e., the variations are not dominated by either short or long baselines.  Furthermore, we checked our result by the more conventional method of restricting our data set to only the longest baselines having $\mathit{uv}$ spacings $>$30 k$\lambda$, calibrating on NRAO 530 and J1744$-$312 separately. The results are consistent with those of the analysis method used here. 

To characterize the flux variations, we subject the real residual 
visibilities of SgrA$^*$ to periodic analysis by deriving a Lomb-Scargle 
power spectrum (Scargle 1982), designed for unevenly 
sampled time-series data. Our high 
sampling rate and daily coverage enable us to explore variations on 
time scales of minutes to hours. Benlloch et al. (2001) illustrate the importance of significance estimation when characterizing time variations in AGN. Thus, to estimate the significance of our power spectra, we resort to Monte Carlo simulation.  Following an algorithm for generating power law noise (Timmer \& K\"onig 1995), we produced 5000 artificial red-noise light curves having the same statistical properties (mean and variance) and time sampling of the original data for each day. The artificial curves have power spectra that obey a power law  with $P(f)\propto{f^{-\beta}}$. Figure 4 is the average of all intra-day 
power spectra of the data (solid curve) and of all simulated light curves with $\beta$ values of 1.0, 1.5, and 2.0. Figure 4 shows that a power spectrum with $\beta=1.0$ is a good fit to the data.  Thus, for each frequency bin, we superimpose on the power spectra in Figure 3 (column 3) the resulting 90\% and 99\% upper confidence level envelopes derived from 5000 simulated power spectra using $\beta=1.0$. 

Figures 3 \& 4 show persistent power on time scales of a few hours, both in most of the individual days and in the average power 
spectrum.  Power at long periods on some days (4--6 hours) is likely to represent an overall flux trend for those days. The rise and decay times of the flux residuals generally occur on a timescale of $\sim$1--2 hours, comparable to the timescales observed by previous experiments (Miyazaki et al. 2004 and Bower et al. 2002). It is unlikely that variable linear polarization or variable sampling of a constant, linearly polarized source through the fixed, linear feeds on the OVRO millimeter array can explain the apparent variations in Figure 3, since the linear polarization of SgrA$^*$ at 3 mm has been constrained to $\lesssim$1\% (Bower et al. 2003). On several days, notably May 27 and 28, there is strong indication of power on time scales of 1.5--3 hours, some of which appears significant at more than the 99\% confidence level derived from our Monte Carlo simulations. This shows up as a peak in the average power spectrum which rises above the best fitting red-noise power spectrum with $\beta$=1 (Figure 4). Finally, we note a suggestion of a correlation between the amplitude of few-hour variability and the average flux on a given day (Fig 2); May 27 and 28, in particular, show the highest mean flux densities. This potential trend clearly needs confirmation. 

\section{DISCUSSION}

The peak at $\sim$2.5--3 hours in the average power spectrum, if confirmed, would suggest that there be a characteristic time scale associated with the 3-mm emission. If we assume that this power spectral peak is due to the dynamical modulation of the emitting gas, we can use the 
black hole mass to compute a dynamical radius of $\sim{0.8}$ AU 
($10R_{s}$). The wavelength dependence of the intrinsic size of SgrA$^*$ taken from VLBA 
closure-amplitude imaging (Bower et al.\ 2004) suggests a major axis 
size of 6$\pm5R_{s}$ at 3 mm, which is consistent with 
the dynamical radius implied by the variability time scale.  We speculate that the absence of significant variations on time scales less than 1 hour is owed to 
the fact that orbital radii with dynamical times less than this are enshrouded in the 
optically thick regime of the accretion flow at 3 mm. 
This would also constrain models in which short-term radio variability originate 
from shocks accompanying a jet (Markoff et al. 2001), if variability of this kind has no 
preferred time scale.

Future long-term monitoring is critical to accurately assess and 
characterize the variability of SgrA$^*$ on short time scales. The future 
Atacama Large Millimeter Array (ALMA) will undoubtedly expand our 
capabilities and help broaden our understanding of this intruiging source.

\noindent\\
{\it{Acknowlegements.}} We would like to thank the Caltech millimeter 
group for generously allocating time on the OVRO millimeter array.  JM would like to thank John Carpenter for useful advice during 
data calibration.

\clearpage


\begin{figure}
\caption{CLEANed OVRO 3mm map of SgrA$^*$ and SgrA West . The familiar 
structure of the ``mini-spiral'' is evident. The beam size is 
9.86$\arcsec$ $\times$ 3.45$\arcsec$ with P.A.= $-$6.22$^\circ$. Contour values 
are 0.02, 0.07, 0.12, 0.17, 0.27, 0.37, 0.57 (black) and 0.77, 1.17, 
1.57, 1.97 (white) Jy/beam. The map was constructed from all eight days of 
data and denconvolved using the AIPS task SCMAP which performed 
phase-only self calibration and CLEANing combined.}\label{fig1}
\end{figure}

\begin{figure}
\caption{Day-averaged flux residuals of SgrA$^*$ on each day of the campaign. The uncertainties 
are given by the standard error of mean of the flux residuals which includes both the uncertainties due to the relative calibrations and noise as well as the true intra-day flux density variations. The absoulute flux on each day is given by 1.7$\pm$0.2 Jy added to the plotted residuals.}\label{fig2}
\end{figure}

\begin{figure}
\caption{Flux residuals for SgrA$^*$ (column 1) and J1744$-$312 (column 
2) with power spectra for Sgr A$^*$ (column 3). The uncertainty for each flux residual represents the relative error and is 
calculated from the RMS variations in J1744$-$312 within $\pm$8 
minutes of each data point. The Lomb power spectra of the data (solid curves) are accompanied by significance levels of 90\% (dotted) and 99\% (dashed) determined using 5000 Monte Carlo simulated light curves with a $f^{-1}$ power spectrum.}\label{fig3}
\end{figure}

\begin{figure}
\caption{Average of all intra-day power spectra over the entire campaign with SgrA$^*$ (solid curve) and  all simulated red-noise light curves with power spectral index $\beta=1.0$ (dashed curve), $\beta=1.5$ (dashed dotted curve) and $\beta=2.0$ (dashed double-dotted curve).}\label{fig4}
\end{figure}

\begin{figure}
\plotone{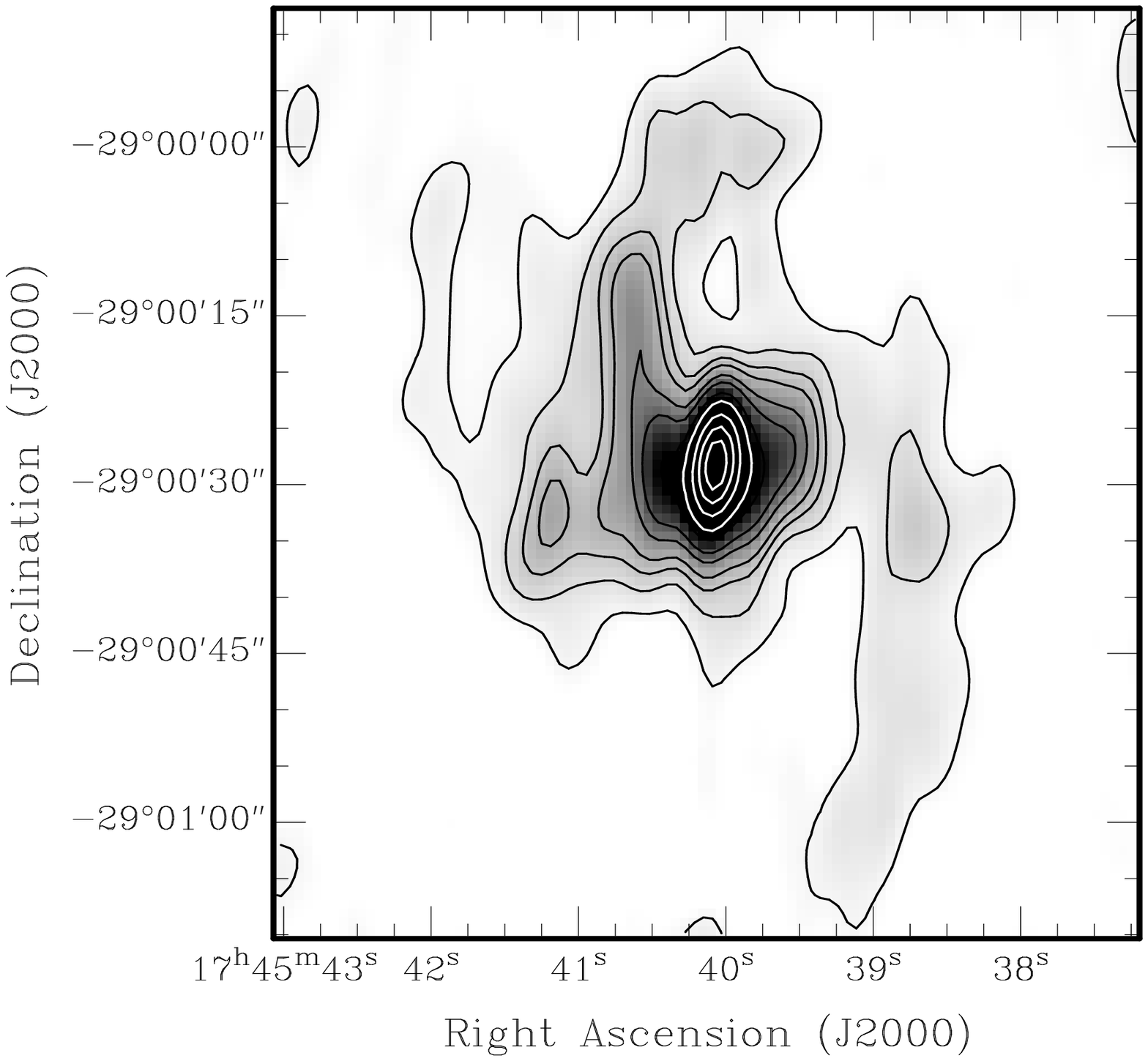}
\end{figure}

\begin{figure}
\plotone{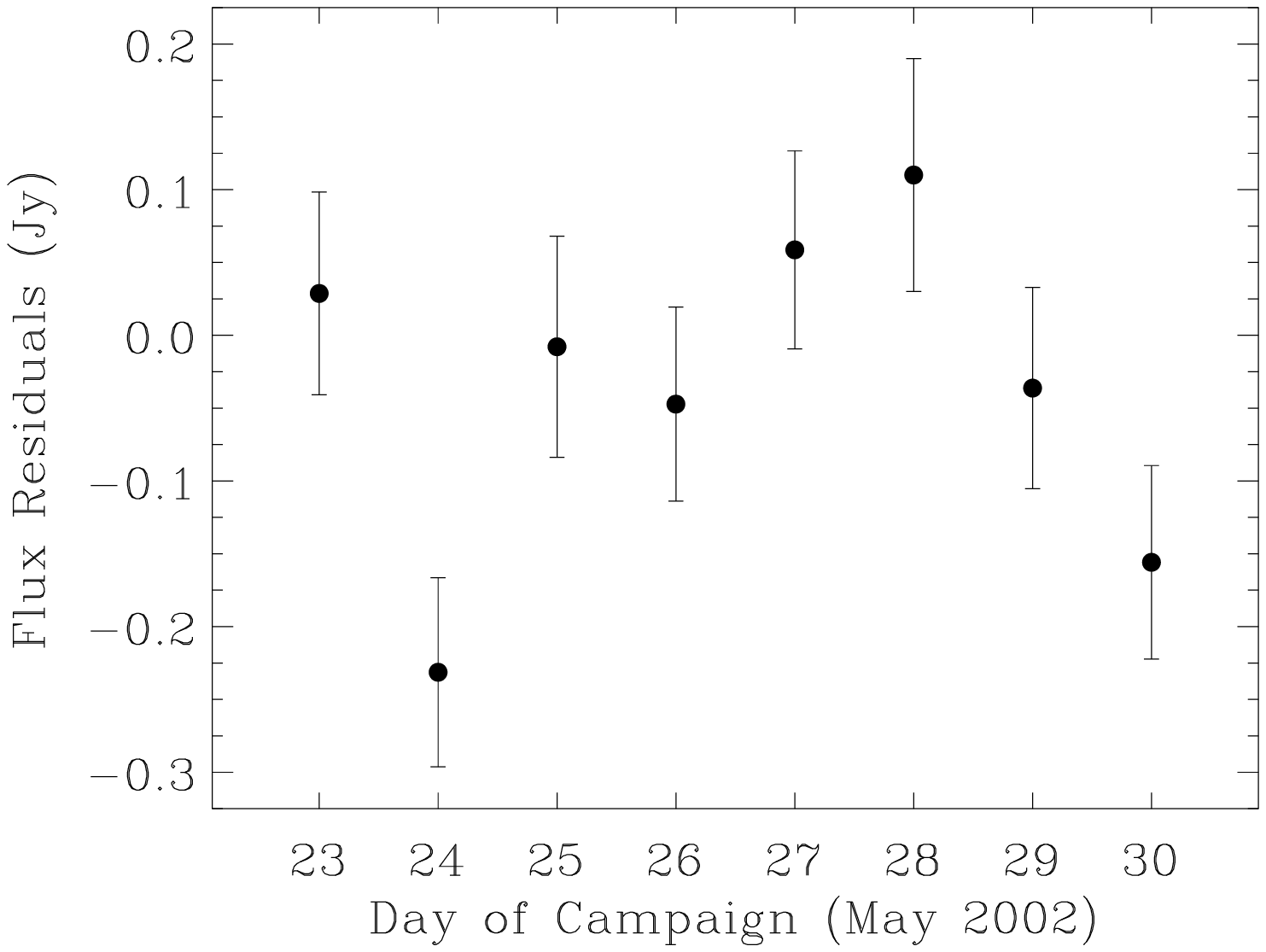}
\end{figure}

\begin{figure}
\plotone{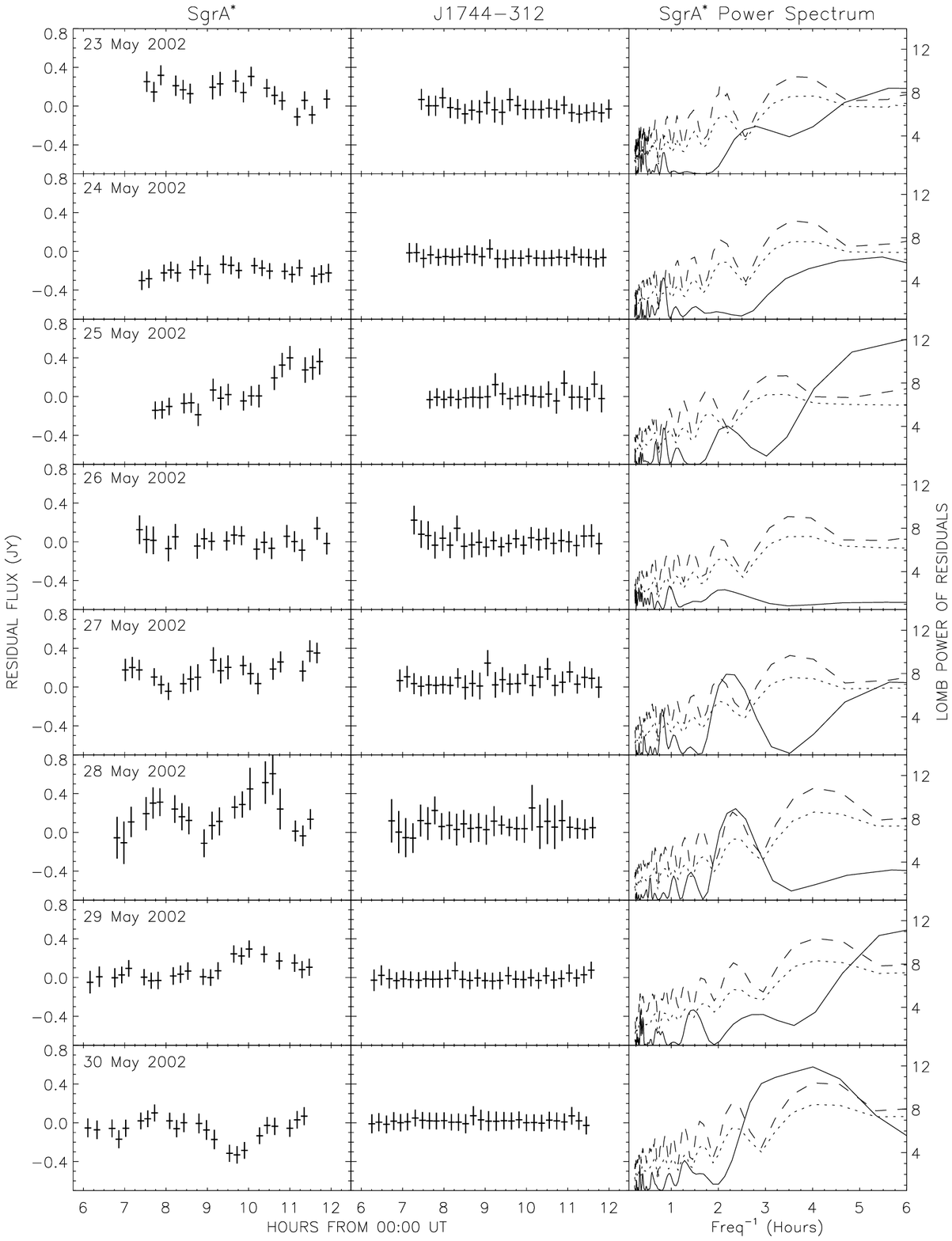}
\end{figure}

\begin{figure}
\plotone{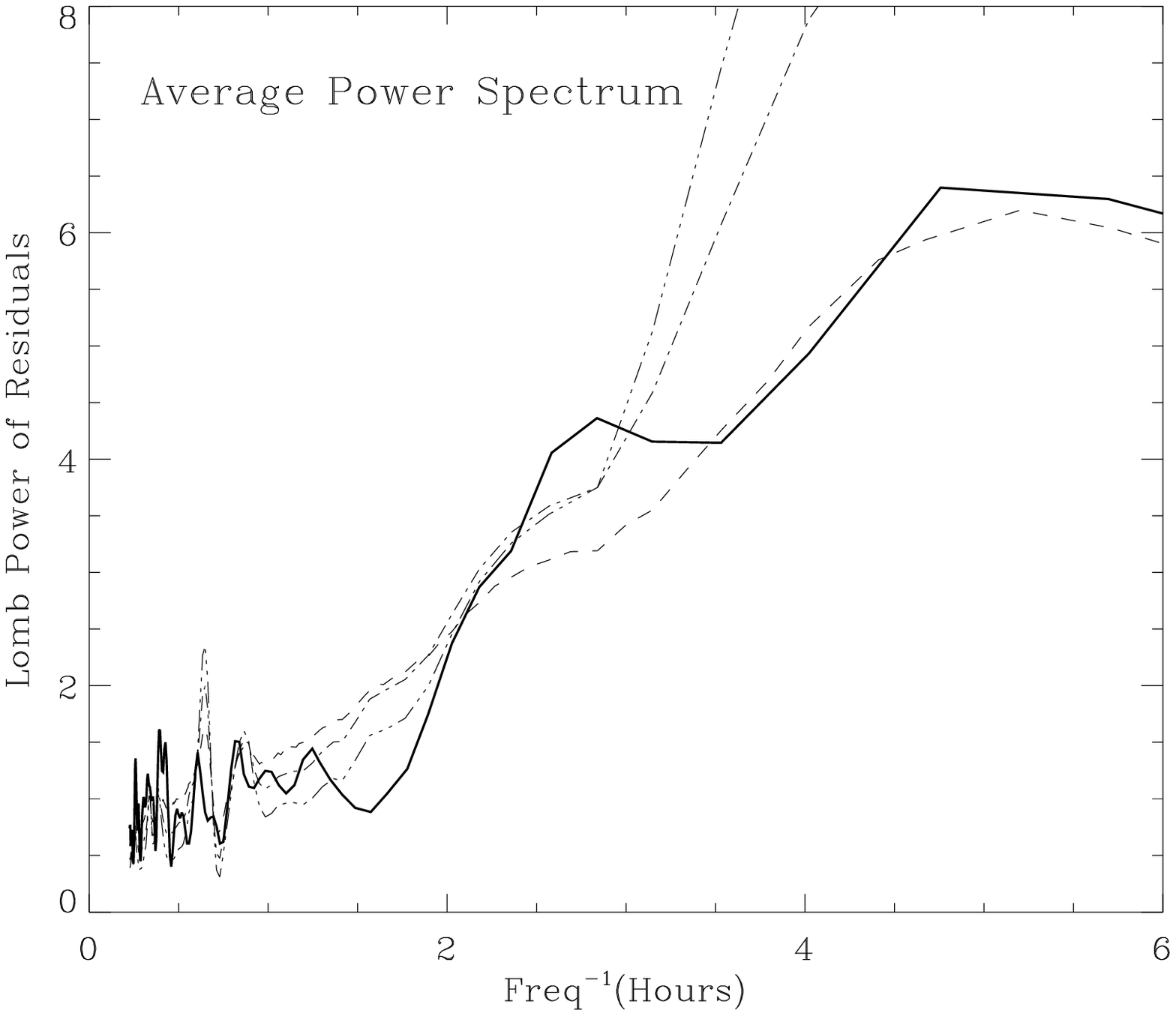}
\end{figure}

\end{document}